\documentclass[twocolumn,epsfig,aps]{revtex4}
\usepackage{epsfig}
\usepackage{amsmath}
\usepackage{amssymb}

\begin{document}
\textheight 235mm

\title{\Large Theoretical study of the thermal behavior of free and alumina-supported Fe-C nanoparticles }
\author{Aiqin Jiang$^1$, Neha Awasthi$^1$, Aleksey N. Kolmogorov$^1$, Wahyu Setyawan$^1$, \\Anders B\"orjesson$^2$, Kim Bolton$^2$, Avetik R. Harutyunyan$^3$ and Stefano Curtarolo$^{1,4}$}
\affiliation{
$^1${\small Department of Mechanical Engineering and Materials Science, Duke University, Durham, NC 27708,}\\
$^2${\small University College of Boraas, SE-501 90, Boraas, and Physics Department, G\"oteborg University, SE-412 96, G\"oteburg, Sweden,}\\
$^3${\small Honda Research Institute USA Inc., 1381 Kinnear Road, Columbus, OH 43212,}\\
$^4$corresponding author: stefano@duke.edu 
}

\date{\today}


\begin{abstract}

The thermal behavior of free and alumina-supported iron-carbon
nanoparticles is investigated via molecular dynamics simulations, in which
the effect of the substrate is treated with a simple Morse potential
fitted to {\it ab initio} data. We observe that the presence of the
substrate raises the melting temperature of medium and large
Fe$_{1-x}$C$_x$ nanoparticles ($x=0-0.16$, $N=80-1000$, non-magic numbers)
by 40-60 K; it also plays an important role in defining the ground state
of smaller Fe nanoparticles ($N=50-80$). The main focus of our study is
the investigation of Fe-C phase diagrams as a function of the nanoparticle
size. We find that as the cluster size decreases in the
1.1-1.6-nm-diameter range the eutectic point shifts significantly not only
toward lower temperatures, as expected from the Gibbs-Thomson law, but
also toward lower concentrations of C. The strong dependence of the
maximum C solubility on the Fe-C cluster size may have important
implications for the catalytic growth of carbon nanotubes by chemical vapor deposition.

\end{abstract}

\maketitle


\section{Introduction}
\label{section.introduction}

Catalytic chemical vapor deposition (CVD) is a widely used method for
the production of carbon nanotubes (CNTs) by decomposition of
hydrocarbons (such as CH$_4$, C$_2$H$_2$ etc.) or carbon monoxide on
supported metal catalysts (Fe, Ni, Co, FeMo, etc.)
\cite{CVD,Ducati,Geohegan,Lee,Avetik5}. Despite numerous studies, the
growth mechanism of nanotubes is still not well understood. Among the
most studied factors that control the growth process are the kinetics
of carbon transport
\cite{Ducati,Geohegan,Lee,Avetik5,Helveg1,Helveg2,Hofmann1,Hofmann2,Hofmann3},
the thermodynamics of the catalyst particles
\cite{Avetik1,Avetik2,Ding1,Ding2,Ding3}, and their interaction with
substrates (oxides or zeolites)\cite{Seidel,Fonseca,Hernadi1}.

The vapor-liquid-solid (VLS) model \cite{VLS} for the CNT growth by
CVD implies that the catalytic particle should be in a liquid state which
allows rapid diffusion of carbon atoms throughout the particle.  The
bulk diffusion of carbon through the metal nanoparticle is driven by
concentration gradients \cite{Audier}, and it is considered to be the
rate-limiting step in the growth of filaments or carbonaceous deposits
\cite{Baker72,Baker75}.  The activation energies ($\sim$1.2-1.8 eV)
measured for thermal CVD growth of nanofibers or nanotubes are
consistent with those for the carbon diffusion through the
corresponding metals\cite{Baker75,Baker78,Ducati,Geohegan,Lee}, hence
further supporting the bulk diffusion VLS model. Another mechanism,
the surface-mediated carbon transport model, has been proposed
\cite{Helveg1,Helveg2}. The low temperature nanotube synthesis by
plasma-enhanced CVD \cite{Hofmann1,Hofmann2,Hofmann3,Chen,Li,Seidel}
implies that the catalyst could be in a solid state. However, the
temperature of the active nanoparticles is extremely difficult to
measure during the growth process.  In fact, the bombardment of
energetic species in plasma and the exothermic dehydrogenation
reactions of hydrocarbon can increase the local temperature of the
catalyst, even if the substrate is kept in thermal equilibrium.
This phenomenon can promote the surface melting of the
nanoparticle and, concurrently, facilitate the growth of the nanotube
\cite{Ding1,Ding2}.

Because the overall catalytic capability of nanoparticles strongly
depends on whether they are in the liquid or solid states, their
thermal behavior has been extensively investigated with experimental
\cite{ex1,ex2,ex3,ex4,ex5,Avetik1,Avetik2,Kriv,Parm} and theoretical
means
\cite{th1,th2,th3,Ding1,Ding2,Michael,Sank1,Aguado,Rossi,Rodri}. Under
CVD experimental conditions, the melting temperatures of catalytic
particles are strongly reduced because of the dissolved carbon ({\it
liquidus} and {\it solidus} slopes in the metal-C phase diagram) and
the relatively high surface energy with respect to the bulk materials
(Gibbs-Thomson phenomenon). Unusually low melting points of
600-700$^\circ$C have been observed for oversaturated solutions of
carbon (up to 50 at.\% at 700$^\circ$C) in Fe, Ni, and Co metals
\cite{Kriv}.  In these cases, the fluidization of the metal catalytic
particles at low or moderate temperature was attributed to the
creation of highly dispersed unstable solutions oversaturated with
carbon, with concentrations well above the limit of the stable carbide
\cite{Parm,Parm_note}.

With so many factors influencing the catalytic growth of CNT, the role
of the substrate on the thermal properties of the particle is often
overlooked. In fact, there are experimental indications that the
presence of substrate could be an important factor in the
thermodynamics of the particle and in regulating the growth of
nanotubes \cite{Seidel,Fonseca,Hernadi1}. Most of the theoretical
research has been focused on the melting behavior of free-standing
pure \cite{th1,th2,th3, Ding2} or bimetallic
\cite{Michael,Sank1,Aguado,Rossi,Rodri} clusters. A phase diagram of
free-standing Fe-C clusters of fixed size ($\sim 2.4$ nm) has been
recently calculated\cite{Ding1,Ding_note}.  There have been
investigations performed on supported catalysts
\cite{Ding3,Sank1,Huang,Antone}, showing that the cluster-substrate
interaction strongly affects the melting temperature of particles and
thus could influence the nanotube growth rate.  A more general
picture, capturing the particle-substrate interaction and size effects
on the phase diagram, is still lacking and is the main subject of our
study.

In this paper we investigate thermal behaviors for free and
alumina-supported Fe-C nanoparticles with molecular dynamics (MD)
simulations. In Section II we describe the {\it ab initio} development
of a simple interaction between Fe and Al$_2$O$_3$ which is essential
for the calculations of supported clusters of reasonable sizes. In
Section III, by using MD simulations we explore the thermodynamics of
Fe nanoclusters and the phase diagrams of Fe-C binary nanoparticles
focusing on the effects of cluster's size and interfacial interactions
on the thermal properties (Fe-C up to $\sim$16\% carbon
concentration). In the same Section we show that not only the eutectic
temperature but also the eutectic composition are size- and
substrate-dependent. Section IV is devoted to the exploration of the
peculiar characteristics of very small clusters.  Conclusions are
described in Section V.


\section{The Fe-Al$_2$O$_3$ potential}
\label{section.potential}

In this section we model the many-body Fe-Al$_2$O$_3$ interaction with
a simple classical potential. If the calculation of this interaction
energy involved the summation over all substrate atoms (e.g. using
modified charge transfer potential \cite{Streitz,Zhou}) the simulation
would be computationally very expensive (a 200-atom nanoparticle would
require a 1000-atom substrate patch). Ideally, a suitable potential
for MD simulations would depend on three integral variables, or even
one (the distance from an iron atom to the surface) if the potential
corrugation is small. The validity of such simplification critically
depends on how weak the Fe-Al$_2$O$_3$ interaction is compared to
those of Fe-Fe and Al$_2$O$_3$-Al$_2$O$_3$.

There are three possible terminations for (0001) $\alpha$-Al$_2$O$_3$
surfaces\cite{Verdo,Ahn, Toofan,Guen}: stoichiometric Al-terminated,
Al-Al-terminated (two top Al layers), and O-terminated. These are
depicted in Figure \ref{label_fig01}(a).  Theoretical calculations
have predicted that the most stable surface is the stoichiometric one,
terminated by a single layer of Al \cite{Verdo}.  This type of surface
is also believed to be the most often observed under ultra high vacuum
conditions \cite{Ahn,Toofan,Guen}.  Therefore in this work we will
restrict our analysis of Fe over Al$_2$O$_3$ using only the stable
stoichiometric Al-termination.
\subsection{Details of Calculation}
\label{section.potential method}
\begin{figure}[htb]
  \begin{center}
    \centerline{\epsfig{file=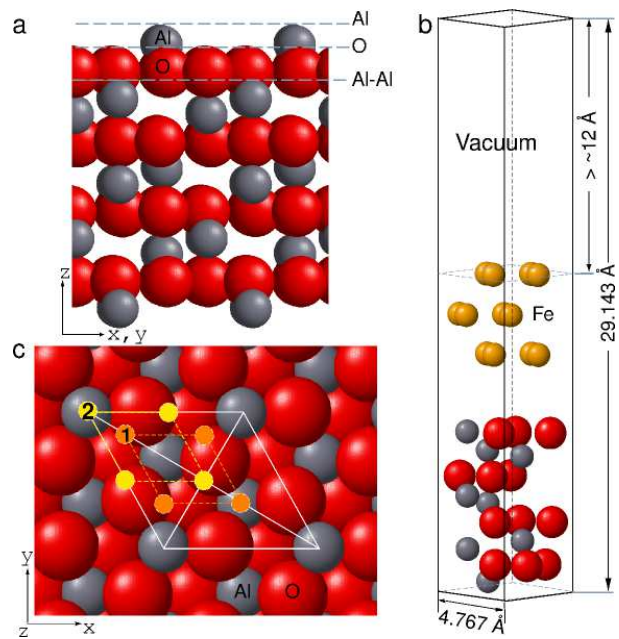,width=80mm,clip=}}
    \caption{\small (color online). 
      (a) Possible surface terminations of $\alpha$-Al$_2$O$_3$ (side view).
      (b) Schematic of the hexagonal unit cell. (c) The two different high symmetry
positions ``1'' and ``2'' of the inner layer of adsorbed iron (top view), each Fe layer has four atoms.
    }
    \label{label_fig01}
  \end{center}
\end{figure}

To develop the interaction between Fe clusters and Al$_2$O$_3$
surfaces, we perform density functional theory {\it ab initio}
calculations with Vienna Ab-Initio Simulation Package {\small VASP}
\cite{Kresse1,Kresse2} with projector augmented waves (PAW)
\cite{PAW,Kresse3} and exchange-correlation functionals as
parameterized by Perdew, Burke, and Ernzerhof (PBE) \cite{PBE} for the
generalized gradient approximation (GGA).  Simulations are carried out
at zero temperature, with spin polarization and without zero-point
motion.  We use an energy cutoff of 500 eV and a $6\times6\times1$
${\bf k}$-point Monkhorst-Pack mesh \cite{MONKHORST_PACK}.  The force
tolerance for structural relaxation is set to 0.1 eV/\AA.  The unit
cell is hexagonal with lattice parameters $a=4.767$ and $c=29.143$
\AA.  The length of the lattice vector normal to surface is kept large
enough to minimize the neighboring supercell interaction. Vertically,
there is at least 12 \AA\ of empty space.  In addition, a dipole layer
is applied in {\it z}-direction to minimize electrostatic effects.  A
schematic of the unit cell is shown in Figure \ref{label_fig01}(b).
\vspace{-8mm}
\subsection{Results and Discussion}
\label{section.potential discussion}
The calculated lattice parameters of the bulk alumina (space-group
\# 167) with a hexagonal unit cell ({\it hR30}), $a=4.767$ \AA\ and
$c=12.994$ \AA, are in good agreement with the experimental values,
$a_{exp}=4.742$ \AA\ and $c_{exp}=12.919$ \AA\, \cite{Amour,PP}.  Our
$\alpha$-Al$_2$O$_3$ slab consists of four oxygen layers as shown in
Figure \ref{label_fig01}(b).  The structure is relaxed from the bulk
$\alpha$-Al$_2$O$_3$ configuration while keeping one oxygen and one
aluminum layers at the bottom frozen. The surface undergoes considerable relaxations:
the interlayer spacings for the top four layers become $d=0.117,
0.899, 1.019$ and 0.264 \AA, differing from the corresponding bulk
values by -87.3\%, 7.8\%, -47.1\%, and 22.1 \%, respectively.  These
values compare well with the previous 18 oxygen-layer simulations (-
87.4\%, 3.1\%, -41.7\%, 18.9\%) \cite{Verdo}, suggesting that an
alumina slab with four layers of oxygen is thick enough to be a
realistic model of the Al$_2$O$_3$ surface.  On the top of the
relaxed $\alpha$-Al$_2$O$_3$ substrate we place a few close-packed
layers of iron while keeping the in-plane lattice vectors of Fe
identical to the bulk value of alumina.  Because of the fortunate
match between the natural Fe and the $\alpha$-Al$_2$O$_3$ surface
lattice spacings (within 3\%), the effect of the interface strain
energy on the total binding is insignificant.  Close-packed iron layers
can be put in different ways on top of alumina.  Thus, we consider
the two high symmetry positions labeled as ``1'' and ``2'' in Figure
\ref{label_fig01}(c).

\begin{figure}[htb]
  \begin{center}
    \centerline{\epsfig{file=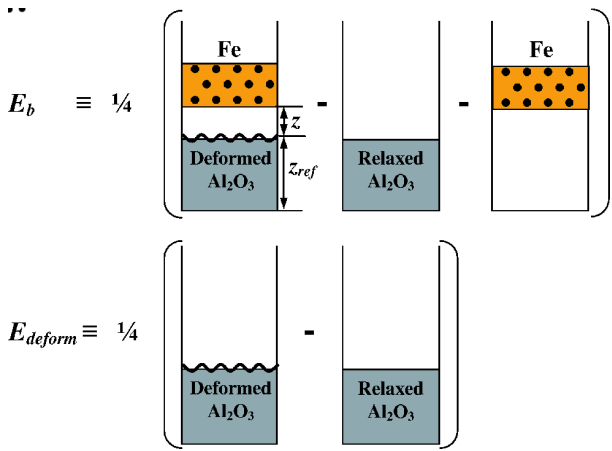,width=80mm,clip=}}
    \caption{\small (color online).
      The definition of the binding and deformation energies.
    }
    \label{label_fig02}
    \vspace{-6mm}
  \end{center}
\end{figure}

The binding energy between one Fe column (one Fe per layer) and
Al$_2$O$_3$ is defined as
\begin{equation}
  E_b \equiv \frac{1}{4}\left[E_{Fe-Al_2O_3}-E_{relaxed[Al_2O_3]}-E_{Fe}\right],
  \label{label_equation_Eb_definition}
\end{equation} 
where $E_{Fe-Al_2O_3}$ is the total energy of the optimized system,
$E_{relaxed[Al_2O_3]}$ and $E_{Fe}$ are the energies of the relaxed
Al$_2$O$_3$ slab and Fe layers, respectively (note that the factor $1/4$
appears because there are four Fe per adsorbed layer).  The
presence of adsorbed Fe strongly modifies the alumina surface, and the
deformation of the surface may affect the total binding considerably.
To evaluate this effect, we define the substrate deformation energy as
\begin{equation}
  E_{deform} \equiv \frac{1}{4}\left[E_{deformed[Al_2O_3]}-E_{relaxed[Al_2O_3]}\right],
  \label{label_equation_Edeform_definition}
\end{equation} 
where $E_{deformed[Al_2O_3]}$ is the energy of an artificially isolated
alumina slab with the same geometry as that of the optimized
Fe-Al$_2$O$_3$ system but without adsorbed iron.  Figure
\ref{label_fig02} shows a schematic of the definitions of the two
energies, $E_b$ and $E_{deform}$.  In our calculations only the
surface is allowed to relax: iron atoms are kept at their ideal
positions because we assume that a reliable Fe-Fe interaction
(e.g. Born-Mayer potential \cite{Guillope,Stanek}) will be able to
properly describe their energetics.

To determine how many layers of Fe must be included in the
calculations, we evaluate the Fe-Al$_2$O$_3$ interaction for systems
with different number of Fe layers situated at the same distance above the alumina
surface.  As summarized in Table \ref{label_table_1}, the binding
energy converges reasonably well once the iron film contains three or
more layers.  Therefore, we limit our calculations to the absorption
of only three Fe layers.
In addition, due to the corrugation of the surface, the binding energy
depends on where the adsorbants are positioned in the {\it x-y} plane.
Figure \ref{label_fig01}(c) shows the two high symmetry configurations
of the inner layer of adsorbed film.  The corrugation, calculated as
the difference between the binding energies of the most and least
stable configurations (position ``1'' and ``2'' in Figure
\ref{label_fig01}(c), respectively), is less than 15\% of the total
energy.  Such little sensitivity to the lateral position of iron
allows us to consider the surface essentially flat.

\begin{table}[hbt]
  \begin{center}
    \begin{tabular}{c|c}
      \hline\hline
      Number of Fe layers           &     $E_b$ (meV) \\
      (4 Fe per layer)            &       \\
      \hline
      One                           &    -666  \\
      Two                           &    -258  \\
      Three                         &    -160  \\
      Four                          &    -159  \\
      Five                          &    -146  \\  
      \hline\hline
    \end{tabular}
    \caption{ \small The dependence of the binding energy on the thickness of 
      close-packed Fe structure with the closest layer located in position ``1'' of Figure \ref{label_fig01}(c).}
  \label{label_table_1}
  \end{center}
\end{table}
\vspace{-2mm}

\begin{figure}[htb]
  \begin{center}
    \centerline{\epsfig{file=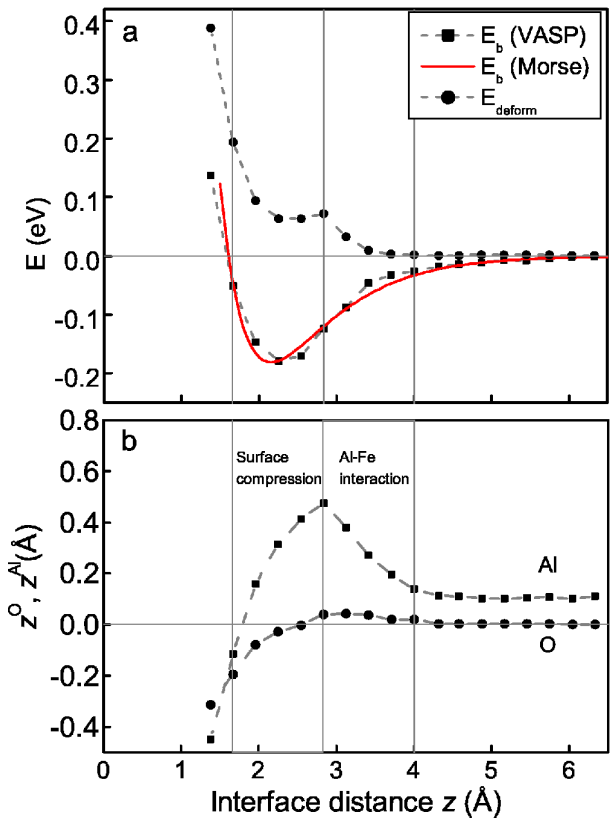,width=80mm,clip=}}
    \caption{\small (color online).
      (a) The binding energy of Fe-Al$_2$O$_3$  ({\it ab initio} values with fitted Morse potential)
      and deformation energy of the Al$_2$O$_3$ slab as functions of the distance from the surface.
      (b) Vertical coordinates of the outer Al and O layers of Al$_2$O$_3$
      as function of the distance from the surface.
    }
    \label{label_fig03}
    \vspace{-6mm}
  \end{center}
\end{figure}

Figure \ref{label_fig03}(a) shows the binding ($E_b$) and deformation
($E_{deform}$) energies as a function of the distance $z$ from the
surface.  Since the Al$_2$O$_3$ slab experiences surface
rearrangement, $z$ must be defined with respect to a fixed reference,
in our case, the bottom of the unit cell (see Figure
\ref{label_fig02}).  For convenience, we use a constant shift
$z_{ref}$ to have the minimum of the interaction energy at a
reasonable distance of 2.25 \AA.  The strong contribution of $E_{deform}$
to $E_b$, shown in Figure \ref{label_fig03}(a), is caused by the
considerable rearrangement of the surface atoms to accommodate the
absorbed iron. The origin of the non-monotonic variation of the
deformation energy is clarified in Fig. \ref{label_fig03}(b), in which
we show the vertical shifts of the outer Al and O layers, $z^{Al}$ and
$z^O$, with respect to the O position when Fe is not present
($z\rightarrow\infty$). The outer positively charged Al layer first
rises above the substrate to be closer to the approaching Fe film and
accept more charge (as has been shown previously for the Zr-terminated
Ni/ZrO$_2$ system\cite{Matsunaga}, the charge transfer between the
outer Zr and Ni, as well as the accumulation of charge in the gap
between them, is significant).  The maximum rise for the Al layer
happens at the interlayer Al-Fe distance of 2.7 \AA ($z$=2.8 \AA\ in
Fig. \ref{label_fig03}(b)), which corresponds to the kink in the
deformation energy curve. After that point, both Al and O layers are
pushed downwards. For $z<1.5$ \AA, one can also expect noticeable
deformation of the Fe film, and the decomposition of the total Fe
binding into the independent Fe-Fe and Fe-Al$_2$O$_3$ contributions may
not be accurate anymore.


The {\it ab initio} binding energy can be decomposed as a sum of contributions from each atom in the Fe ``column''
(assuming that the many-body Fe-Fe effects are appropriately described 
by an external interaction):
\vspace{-2mm}
\begin{equation}
  E_{b}=\sum_{i=1}^3 E_{i}
  \label{label_equation_01}
\end{equation}
where the energy per atom $E_{i}$ is taken to be a simple Morse potential\cite{Morse} as:
\begin{equation}
  E_{i}=D \left\{\exp{\left[-2\alpha\left(z_{i}-z_{0}\right)\right]}-2\exp{\left[-\alpha\left(z_{i}-z_{0}\right)\right]}\right\}
  \label{label_equation_02}
\end{equation}
Fitting gives $D=153$ meV, $\alpha=1.268$ \AA$^{-1}$, and $z_0=2.219$ \AA.
Figure \ref{label_fig03}(a) shows the fit of $E_b$ 
calculated as a sum of three Morse interactions.

The strength of the Fe-Al$_2$O$_3$ interaction (153 meV) amounts to
only $\sim$10\% of the Fe-Fe interlayer binding energy
($\sim$1.24-1.42 eV, depending on the thickness of the Fe film), which
implies that the latter is not significantly affected by the presence
of the substrate.  Hence, the assumption of decomposing the total Fe
binding into the separate Fe-Fe and Fe-Al$_2$O$_3$ interactions is
justified {\it a posteriori}.  Our simple Fe-Al$_2$O$_3$ Morse interaction
naturally incorporates the whole alumina surface deformation, since
the binding energy has been defined with respect to an ideal
Al$_2$O$_3$ slab and an isolated Fe film, as described in equation
(\ref{label_equation_Eb_definition}) and Figure \ref{label_fig02}.
Within this framework, MD simulations on supported particles can be
efficiently performed with reasonable accuracy by considering the $z$
component of the particle interaction with the flat substrate.

\def\EbFF{$E_b$[Fe$_{11}$C$_1$-Al$_2$O$_3$]}
\def\EbFC{$E_b$[Fe$_{12}$-Al$_2$O$_3$]}

As we are interested in the properties of binary Fe$_{1-x}$C$_x$
nanoparticles, we need to evaluate the interaction between carbon
surrounded by iron and the alumina substrate
(Fe-embedded-C-Al$_2$O$_3$ interaction). Starting from three layers of
adsorbed Fe, we substitute one Fe with one C atom bringing the
concentration to $x\sim$8.3\%. Depending on the position in which the
substitution takes place, the binding energy \EbFF\, fluctuates while
remaining similar to \EbFC, which suggests some importance of
many-body effects in the Fe-embedded-C-Al$_2$O$_3$ interaction.
Because the substrate represents only a small perturbation to the
total binding of the nanoparticle and the concentration of C in our
simulations is small, we choose to treat the C-Al$_2$O$_3$ and Fe-Al$_2$O$_3$
interactions in the same way, i.e. we use the Fe-Al$_2$O$_3$ Morse
potential for both atom types. The validity of this approximation has
been addressed by performing MD simulations, similar to those
described in Section \ref{section.MD}, but here we vary the Morse
parameters for the C-Al$_2$O$_3$ interaction.  Our tests for supported
Fe$_{300}$C$_{30}$ nanoparticles reveal that the thermodynamics of the
systems is not very sensitive to the particular value of the
C-Al$_2$O$_3$ binding.  For example, increasing or decreasing the
strength of the C-Al$_2$O$_3$ interaction for the Fe$_{300}$C$_{30}$
nanoparticles by a factor of two changes the melting temperature by
1-2 \%, the amount comparable to the statistical error of the melting
temperature determination in our simulations(see Section III). We
conclude that the simple approximation of employing the same
description for the Fe and C interactions with the substrate is
suitable for this study.


\section{Molecular Dynamics}
\label{section.MD}

Among the taxonomy of thermodynamics phenomena for nanoparticles,
melting has been subject of considerable interest.  The
characteristics of the melting process depend on a variety of
parameters such as size \cite{ex1,LL} and shape of the particles
\cite{Wautelet}, concentration of impurities \cite{Mottet}, and
presence of substrates \cite{Huang,Antone}.

In this section we are interested in the effect of size, carbon
presence and Al$_2$O$_3$ substrate on the melting and solidification
of binary Fe-C nanoparticles.  We address these tasks by analyzing,
with classical MD simulations, the liquidus and
solidus lines of their phase diagrams, in function of the
aforementioned parameters.

\subsection{Methods}
\label{section.MDmethod}

\begin{figure}[htb]
  \begin{center}
    \centerline{\epsfig{file=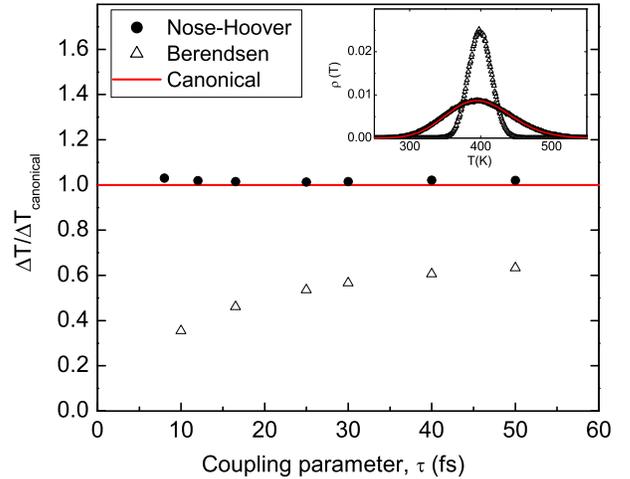,width=90mm,clip=}}
    \caption{\small (color online).
      Comparison of Berendsen and Nos\'e-Hoover thermostat for $N_{Fe}=50$ at 400 K for $\Delta t$ = 1 fs: 
	temperature dispersion for Nos\'e-Hoover is closer to the canonical dispersion. 
        Inset shows the distribution of instantaneous kinetic temperature for Berendsen thermostat for $\Delta t/\tau = 0.1$, 
        Nos\'e-Hoover thermostat for $\tau = 25$ fs, and the canonical distribution.}
    \label{label_figure_04}
    \vspace{-4mm}
  \end{center}
\end{figure}

{\it Molecular Dynamics Simulations.\,\,} MD simulations are carried in
the $NVT$ ensemble using the Verlet algorithm \cite{ver1,ver2} with
a time step $\Delta t$ = 1.0 fs. Of the several methods developed for
controlling the temperature in MD simulations
\cite{Berendsen,Nose1,Nose2,Hoover,Langevin}, Berendsen and Nos\'e-Hoover
thermostats are most commonly used. In Fig. \ref{label_figure_04} we
compare the temperature dispersion $\Delta T$ of the two thermostats
with respect to that of the canonical distribution $\Delta
T_{canonical}$ \cite{Holian} at $T$ = 400 K for small nanoparticles
($N_{Fe}=50$). We observe that the widely used Berendsen thermostat is
more sensitive to the choice of the coupling constant. Inset in Figure
\ref{label_figure_04} shows the distribution of instantaneous kinetic
temperature for Berendsen thermostat for the typical value of $\Delta
t/\tau$ = 0.1 \cite{YHu}, the Nos\'e-Hoover thermostat for $\tau$ = 25
fs, and the canonical distribution at $T =$ 400 K. The Nos\'e-Hoover
thermostat ($\sigma_T\sim$ 47 K) reproduces the canonical
distribution($\sigma_T\sim$ 46 K) much better than the Berendsen
thermostat ($\sigma_T\sim$ 17 K), making it a better choice for our
constant temperature simulations.

%
%
%
%

{\it Interatomic interaction.\,\,} 
Fe-Fe, Fe-C, and C-C interactions are described by
Born-Mayer \cite{Guillope,Stanek}, Johnson \cite{Johnson1,Ding4} and
Lennard-Jones \cite{Girifalco} potentials, respectively. 
These interaction models are discussed in detail elsewhere \cite{Ding4}. 
The Morse potential, introduced in Section \ref{section.potential}, 
is used to model the Fe-Al$_2$O$_3$ interaction; the Fe-embedded-C-Al$_2$O$_3$ 
interaction (C is diffused in Fe) is taken to be identical to Fe-Al$_2$O$_3$ 
as discussed before.

{\it Initial configurations.\,\,} To avoid excessive temperature
fluctuations in the MD simulations of the nanoparticles one should
start from the most stable configurations. We search for the best
possible energy minima by randomly arranging atoms in a spherical
nanoparticle, carefully optimizing the positions of iron and carbon
atoms and finally annealing the nanoparticles for $6\times10^6$ MD
iterations (6 ns). The annealing is performed in the following way:
the nanoparticle is first heated to high temperature (from 1000 K to
1400 K depending on the size of the particle) for $0.6\times10^6$
steps, kept at constant temperature for another $0.6\times10^6$
iterations, and finally cooled to 0 K during the remaining
$4.8\times10^6$ MD steps.

{\it Definition and determination of melting temperature.\,\,} In our
work, the melting phenomenon is analyzed by performing several MD
simulations starting at about 300 K below the expected melting point
with temperature increments of 10 K for small ($N<100$) and 20 K for
large clusters (with 5 K upon approaching the transition).  Only the
lowest temperature simulations begin from the annealed initial
structures: the others start from the final configurations (positions,
forces, velocities) of the preceding temperature simulation.  Data
gathering of the energies and other averages are performed over $10^6$
MD steps.

\begin{figure}[htb]
  \begin{center}
    \centerline{\epsfig{file=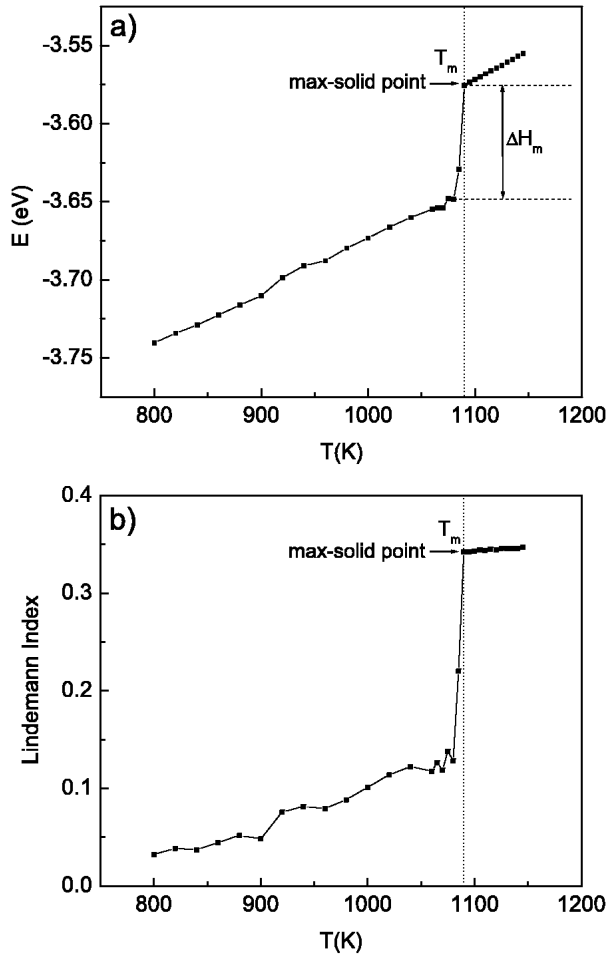,width=90mm,clip=}}
    \vspace{-6mm}
    \caption{\small
      Estimation of the melting temperature $T_m$ as {\it max-solid point}
      for Fe$_{400}$ nanoparticle 
      from the caloric curve (panel a) and from the Lindemann index plot (panel b).}
    \label{label_figure_05}
    \vspace{-6mm} 
  \end{center}
\end{figure}

Several dynamical and structural properties such as total energy,
Lindemann index, diffusion coefficients, and pair correlation functions
can be used to identify phase transitions in nanoparticles
\cite{Alavi,Zhu}.  Here, melting is characterized by the temperature
dependence of the total energy (caloric curve), by the change in the
total energy with time, and by the variation of the Lindemann index
with respect to temperature \cite{Lindemann}. The Lindemann index
$\delta$ represents the root-mean-square relative bond-length fluctuation:
\begin{equation}
  \delta \equiv \frac{2}{N(N-1)} \sum_{i<j} \frac{\sqrt{{\langle r_{ij}^2 \rangle} - {\langle r_{ij} \rangle}^2}}{\langle r_{ij} \rangle},
  \label{label_equation_lindemann}
\end{equation}
where $r_{ij}$ is the distance between atom $i$ and $j$, $N$ is the
number of particles and the average is calculated over an MD run at a
given $T$. The melting point, which defines the temperature at which a
solid becomes liquid, is a macroscopic concept for pure and bulk
systems.  Both finite-size and presence of more than one atomic
species make the melting transition a continuous phenomenon that
occurs over a range of temperatures, $\Delta T_m$, in which solid and
liquid phases coexist with different fractions
\cite{Lupis,McQ,Labastie}. To have a specific value of $T_m$ instead
of a range, we define the melting temperature $T_m$ as the ``{\it
max-solid point}'' which represents the maximum temperature at which
the solid and the liquid phases coexist (the locus of all the
max-solid points is the liquidus). Above $T_m$, no solid phase is
present. Note that within this definition of $T_m$, we also identify
plastic-viscous nanoparticles as ``liquid''
\cite{Iijima-PRL86}. Figure \ref{label_figure_05} shows an example of
$T_m$ calculated from the caloric curve (panel a) and from the
Lindemann index (panel b). Similarly, the {\it min-liquid point} is the
minimum temperature at which the solid and the liquid phases coexist
(the locus of all the min-liquid points is the solidus).  The difference between the
energies of the particle at the max-solid and at the min-liquid points
defines the enthalpy of melting $\Delta H_m$.

\subsection{Melting of nanoparticles}
\label{section.Melting}

\begin{figure}[htb]
  \begin{center}
    \centerline{\epsfig{file=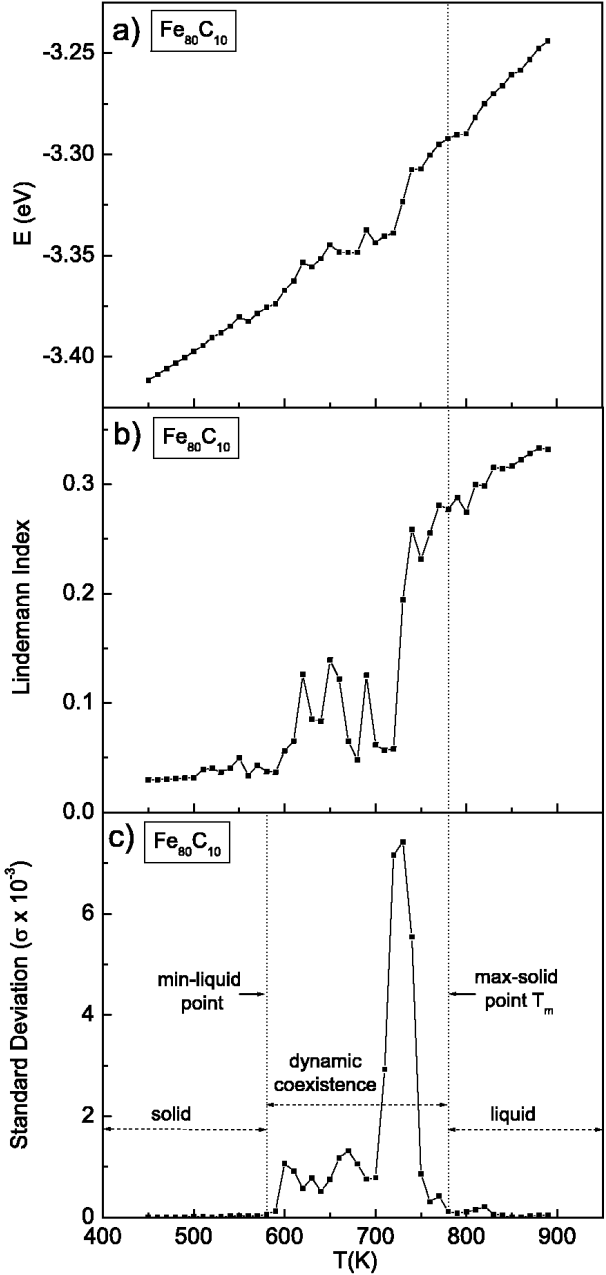,height=180mm,clip=}}
    \vspace{-6mm}
    \caption{\small
      Melting phenomenon for small nanoparticles (N$_{Fe}\sim$80-100 atoms, $d\sim$1 nm):
      a) caloric curve, 
      b) Lindemann index with respect to temperature,
      c) standard deviation $\sigma_{LI}$ of the Lindemann index to identify the max-solid and min-liquid points.
    }
    \label{label_figure_06}
    \vspace{-2mm}
  \end{center}
\end{figure}

With the aforementioned method, we investigate pure nanoparticles of size 
$N_{Fe}=80-1000$ atoms (diameter $d\sim$ 1-3 nm). 
Caloric curves for particles with $N_{Fe}>100$
show small melting intervals ($\Delta T_{m}\lesssim10$ K). 
For smaller clusters (N$_{Fe}\sim 80-100$ atoms, $d\sim$1 nm)
the characterization becomes difficult because 
both the caloric curves and the Lindemann indices 
fluctuate over wide intervals of temperatures 
in which the liquid and solid phases coexist in dynamic equilibrium.
This well known phenomenon, called {\it dynamic coexistence melting}
\cite{Alavi} and shown in Figures \ref{label_figure_06}(a) and
\ref{label_figure_06}(b), is caused by the multitude of different
metastable solid phases present at the nanoscale, which have similar
free energies, similar volumes (at constant
pressure volumes are allowed to change) and different surface
arrangements.
The particles are quasi-plastic 
by continuously changing their state
while alternating metastable configurations and liquid states (bi-stability).
Thus, the observations of thermal properties inside the dynamic
coexistence melting interval $\Delta T_{m}$ are affected by two types
of fluctuations: physical, due to coexistence of phases (cannot be
avoided), and statistical (can be reduced by increasing the total time
of the MD simulation).  In particular, the fluctuations of the
Lindemann index can be captured by analyzing the standard deviation
$\sigma_{LI}$.  Figure \ref{label_figure_06}(c) illustrates the
phenomenon: by plotting $\sigma_{LI}$ versus $T$, we can estimate the
dynamic coexistence melting interval $\Delta T_{m}$, the max-solid and
min-liquid points.  Pure liquid and solid temperature ranges will be
the ones with negligible $\sigma_{LI}$.

\begin{figure}[htb]
  \begin{center}
    \centerline{\epsfig{file=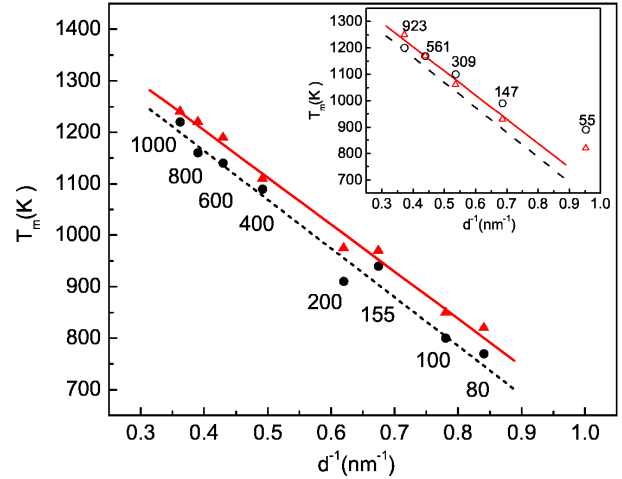,width=80mm,clip=}}
    \vspace{-2mm}
    \caption{\small (color online). Melting temperature versus the
      inverse of particle diameters. Free non-magic sizes ($\bullet$)
      and their linear fit (dashed line); supported non-magic sizes
      ($\blacktriangle$) and their linear fit (solid line); Supported
      clusters have higher melting point due to the decreased
      curvature of the surface\cite{Ding3}. Melting temperatures of
      free magic sizes ($\circ$) and supported magic sizes ($\triangle$)
      clusters were compared with the linear fit lines of
      non-magic sizes (inset). }
    \label{label_figure_07}
    \vspace{-6mm}
  \end{center}
\end{figure}

Figure \ref{label_figure_07} shows $T_m$ of pure Fe nanoparticles in
the whole range of sizes ($N_{Fe}=80-1000$) for free and supported
clusters.  Our results, in agreement with other theoretical
\cite{th1,th2}, computational \cite{Ding1,th3}, and experimental
studies \cite{ex1,ex2,ex3,ex4,ex5}, predict a decrease in the melting
temperature inversely proportional to the cluster diameter \cite{JJ}.
The behavior of $T_m$ can be described by the model based on the
Gibbs-Thomson equation \cite{Lupis,ex4,Hanszen} in function of bulk
melting temperature $T_m^{bulk}$, effective diameter of the particle
$d$, latent heat of melting $\Delta H_{sv}$, and solid-vapor
interfacial energy $\gamma_{sv}$ \cite{Celestini}.


The melting point of bulk Fe, obtained by extrapolating the fit to
$d\rightarrow\infty$ ($T^{bulk}_m(N_{Fe}=\infty)\sim$1416 K) is $\sim
20\%$ below the real one \cite{Massalski}, indicating that our Fe-Fe
Born Mayer many body potential is slightly underbinding.  The
systematic shift in melting temperatures should not affect the main
conclusions of our study since we are interested in the trends of the
liquidus lines in the phase diagrams rather than their precise values.
The supported particles considered here have higher melting point than
that of the free clusters.  In fact, the attractive interaction with
the substrate (Al$_2$O$_3$) induces flattening of the particle and
increases the effective diameter in agreement with previous studies
\cite{Ding3}.

Melting temperatures for magic-size clusters ($N=55, 147, 309, 561,
923$) in our range of simulations are shown in the inset of Figure
\ref{label_figure_07}.  Due to their inherent symmetry (icosahedral or
decahedral for small clusters in our case), free magic-size clusters
(smaller than $N_{Fe}=$ 309) are very stable with melting temperatures
higher than that of the non-magic ones.  This does not apply for
supported clusters.  In fact, the presence of substrate interaction
changes the magic-size sequence: a cluster of magic-size if free, is
not magic if supported (because of the shape and sometimes
structural modifications due to interaction with the attractive
surface; the phenomenon is addressed in Section
(\ref{section.smallclusteres})). In summary, the small supported
magic-size clusters have lower melting point than the corresponding
unsupported ones.

The melting temperature of very small clusters ($N \lesssim 80$)
is a function not only of their size but also of their specific
structure.  It has been shown theoretically that melting temperatures
of clusters with several or tens of atoms can be abnormally high, even
above the corresponding bulk values \cite{AAS,Chuang,Chacko,GAB}.
This phenomenon is related to atomistic processes of structure
isomerization (geometry reconstruction) \cite{Soule,AAS,Chuang,GAB} or
electronic structure change (formation of strong covalent bonds)
\cite{Chacko,Joshi}.  Since the thermodynamics of magic-size
nanoparticles is significantly influenced by their structure, they
experience unusual peculiarities.  In this Section we focus our
analysis only on the non-magic size ones.

\subsection{Phase diagram of free and supported Fe-C nanoparticles}
\label{section.Phase diagram}

\def\cconc{x^C}
{\it Analysis of phase diagrams.\,\,} To understand
how inclusion of carbon atoms influences the thermal behavior of the
catalyst nanoparticles, we determine the melting temperatures as a
function of carbon concentration $\cconc$ ranging from zero to up to
$\sim$ 16 \%. To appropriately model the nanocatalyst in a nanotube
growth process, instead of substituting Fe with C to increase the
concentration of carbon, we add $N_C$ atoms to the Fe particle. Since
$N_{Fe}$ remains constant and $N_C$ increases, the concentration of
carbon is defined as $\cconc\equiv N_{C}/(N_{Fe}+N_{C})$. We plot
the locus of the max-solid points which represent the
liquidus of the Fe-C phase diagram.
In small Fe nanoparticles one might expect the average radius to
increase noticeably with the addition of a few C atoms ($r^3\propto
N$) and hence their melting temperature would also increase
($(T_m^{bulk}-T_m)/ T_m^{bulk} \propto d^{-1}$ ). However, our tests
show that addition of C ($\cconc\lesssim 16\%$) does not significantly
change the volume of the nanoparticle indicating that C behaves as an
interstitial solute in Fe nanoparticles as it does in bulk Fe
\cite{BJLee}.

\begin{figure}[htb]
  \begin{center}
    \centerline{\epsfig{file=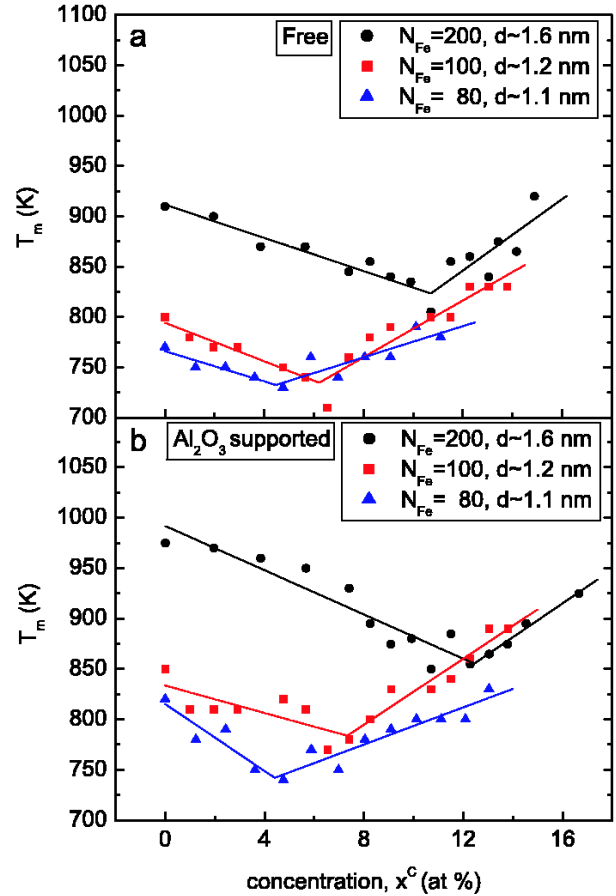,width=90mm,clip=}}
    \vspace{-6mm}
    \caption{\small (color online).
      Fe-C phase diagrams obtained with addition of C (upto $\sim 16\%$) to particles with $N_{Fe}=$80,100, and 200: 
      a) free nanoparticles and b) Al$_2$O$_3$ supported nanoparticles.
    }
    \label{label_figure_08}
    \vspace{-6mm}
  \end{center}
\end{figure}

Figure \ref{label_figure_08} shows phase diagrams ($T_m$ versus
$\cconc$ in atomic \%) for free and Al$_2$O$_3$-supported particles
with $N_{Fe}=$ 80, 100, and 200 based on caloric curve and Lindemann
index analysis. All the data sets show a similar trend in function of
C concentration: $T_m$ decreases almost linearly at low $\cconc$ and
then increases for all the higher $\cconc$ considered.  The exact
functional form is difficult to determine because of the dispersion in
the data, however the observed {\it ``V''-shape} dependence is
consistent with that in the bulk Fe-C phase diagram \cite{Massalski}.
Hence, by using the least square method we approximate the liquidus
with a set of two straight lines, the intersection of which gives the
eutectic point ($\cconc_{eut}$, $T_{eut}$) \cite{LSFit}. This
procedure allows us to estimate this invariant point with an accuracy
of 1 $\%$ and 12 K for $\cconc_{eut}$ and $T_{eut}$, respectively.

We observe that as the particle size is reduced, the eutectic point
for free and supported nanoparticles moves toward lower temperatures
and lower concentrations, indicating that the solubility of C
decreases as well. To the best of our knowledge this phenomenon has
never been reported before. Explanation of its origin will require a
more detailed study of the behavior of dissolved C at various
concentrations (changes in the distribution of C across the particle,
possible formation of stable carbides etc.).
%
%

%

{\it Implications for carbon nanotube growth.\,\,} The ``V''-shape
liquidus feature of Fe-C nanoparticles observed in our simulations
allows for the VLS interpretation of experimental results for
catalytic activity of Fe.  For a given temperature above the eutectic
temperature $T_{eut}$, the dissolution of carbon in a metal catalyst
initially induces the liquefaction of the particle by lowering its
melting temperature \cite{Avetik1,Avetik3}.  As the catalyst becomes
less viscous, the diffusion of carbon in the particle increases. High
catalytic activity has been observed with associated liquefied
particles during the growth of SWNT by CVD method \cite{Avetik3}. Some
experiments have shown presence of liquid-like features and liquid
layers on nanoparticles \cite{Ercolessi,Kofman} before complete
melting. These features would enhance the diffusion of carbon and the
subsequent melting of the particles. In our case, the nanoparticles
are so small that the surface effects on the melting temperature
should be dominant.  In fact, one cannot distinguish between
surface and bulk layers in a particle with 200 atoms of radius $\sim$
0.8 nm, which is approximately three close-packed layers.

According to the data in Fig. \ref{label_figure_08}, if during CVD
experiments the size of the catalyst particle were in a given range
($d\sim 1-2$ nm), the smaller Fe nanoparticles ($d \sim 1$ nm) would
liquefy earlier then the bigger ones ($d \sim $ 2 nm). Consequently,
if liquefaction was a prerequisite for catalytic activity, the small
particles would begin to produce nanotubes earlier.  Experimental
verification of this hypothesis is challenging: very small catalyst
nanoparticles ($d \cong$1 nm) have a fast rate of coalescence during
the reaction and coagulate to form bigger clusters
\cite{Avetik4,Wadhawan}.  Following the VLS model, the growth of very
small nanotubes using metallic catalyst would be possible only if the
reaction temperature were above $T_{eut}$ (to liquefy the particle) and
below the temperature at which particles begin to coalesce.  Hence,
controlling the diameter of nanotubes grown with CVD in the small size
range might be difficult.  So far, the smallest reported nanotube (d =
4 \AA, the most internal tube in a multiwalled nanotube) has been
grown by arc-discharge method without any metallic catalyst
\cite{Qin}.

In the $\cconc > \cconc_{eut}$ region, our results also show that once
the dissolved carbon concentration reaches a point where the
corresponding melting temperature (from the phase diagram) exceeds the
reaction temperature, the particle starts solidifying as a two-phase
system composed of solid carbide and Fe-rich liquid.  This
solidification gradually reduces the average surface mobility of the
catalytic species (Fe) and might affect the catalytic activity of the
nanoparticle. During the process of nanotube growth, conditions such
as sufficient carbon concentration and temperature can induce the
formation of stable carbides. Cementite (Fe$_3$C) and other iron
carbides have been observed in experiments after the reaction of
hydrocarbon or carbon monoxide with iron catalysts
\cite{Sacco,Oberlin,Kock}.  It has been reported that such stable
carbides act as poison by terminating the growth of nanotube
\cite{Gavillet,Avetik2}.

\section{Peculiarities of small free and supported Fe clusters}
\label{section.smallclusteres}

\begin{figure}[b]
  \begin{center}
    \centerline{\epsfig{file=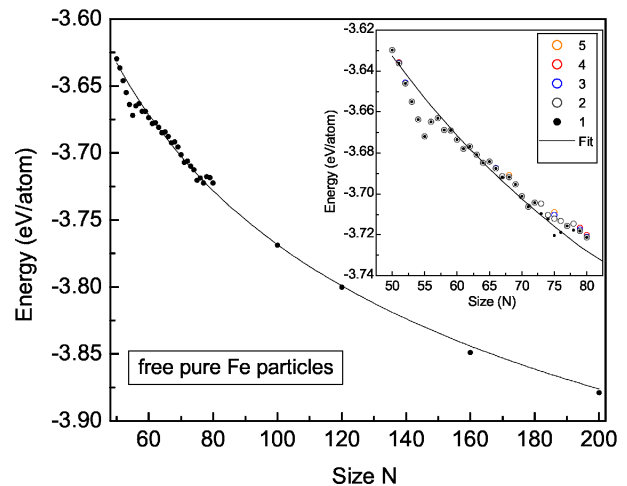,width=80mm,clip=}}
    \caption{\small (color online).
      The minimum energies of free clusters and the
      fitting curve using $E_{ave} = a + b N^{-1/3}$.
      The 5 lowest energies
      are shown in the insert for cluster`s sizes from 50 to 80.
    }
    \vspace{-6mm}
    \label{label_figure_10}
  \end{center}
\end{figure}

In this section we analyze the structural properties of pure small Fe
clusters, free and supported on Al$_2$O$_3$. We find that as the size
decreases Fe clusters assume stable polyhedron
configurations\cite{Wales,Doye}.  The same MD technique described in
the previous section is adapted for searching the stable
configurations of small clusters ($N\sim50-80$). Twenty random
configurations are generated per each size and optimized by
annealing. The whole annealing process contains $10^7$ MD iterations
(10 ns).  Free and supported clusters are heated to 1200 K and 1400 K,
respectively, before being slowly cooled to 0 K.  Among the twenty
annealed structures, we consider the lowest energy configuration as
the global minimum for each size.  For free clusters with $N<72$, the
lowest 5 energies are nearly degenerate (inset in
Fig. \ref{label_figure_10}) which validates our approach.  For bigger
particles the found minima might be sub-optimal as the difficulty of
finding the global minima rapidly increases.

\def\polyfcctrans{polyhedron$\rightarrow$fcc transition\,}

\begin{figure}[t]
  \begin{center}
    \centerline{\epsfig{file=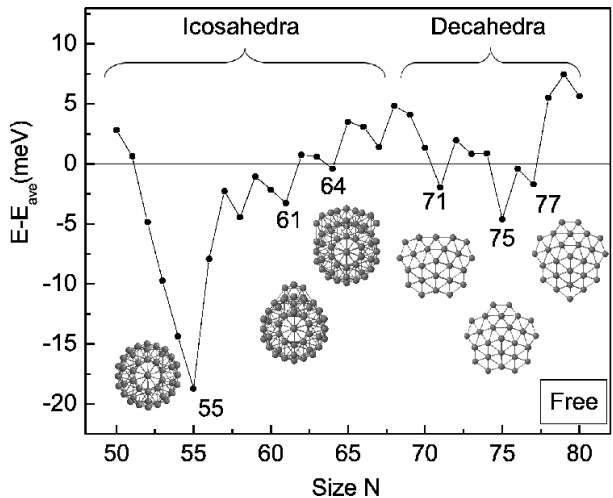,width=80mm,clip=}}
    \vspace{-1mm}
    \caption{\small
      Energies of best minima for free clusters relative to $E_{ave}$, with a selection of stable structures.
      For free clusters, the \polyfcctrans is continuous and occurs around $N\sim 200-500$ (not shown).
    }
    \label{label_figure_11}
    \vspace{-6mm}
  \end{center}
\end{figure}

\begin{figure}[t]
  \begin{center}
    \centerline{\epsfig{file=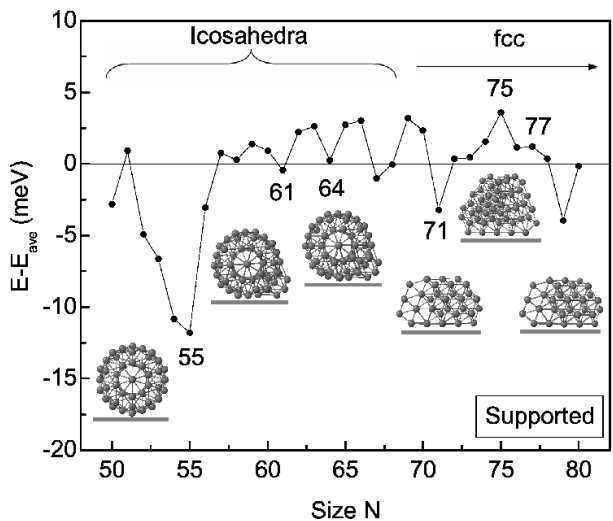,width=80mm,clip=}}
    \vspace{-1mm}
    \caption{\small
      Energies of best minima for supported clusters relative to $E_{ave}$, with a selection of structures.
      For alumina supported clusters, the \polyfcctrans occurs around $N\sim70$.
    }
    \label{label_figure_12}
    \vspace{-6mm}
  \end{center}
\end{figure}

The minimum energies per atom are fitted to the $E_{ave}=a+bN^{-1/3}$
dependence, where, in the case of spherical particles, the parameters
$a$ and $b$ would represent the bulk energy per atom and the
surface-creation destabilization energy cost, respectively
\cite{Lee2,Northby,Xie}. Our fit uses only two parameters $(a, b)$
instead of the four of the more general formula
$E_{ave}=a+bN^{-1/3}+cN^{-2/3}+dN^{-1}$ \cite{Lee2,Northby,Xie} to
avoid overfitting due to the limited number of energy points. We
include the best available minima obtained with the aforementioned
procedure for bigger ($N =$ 100, 120, 160 and 200) clusters to ensure
the correct assymptotic behavior for the nanoparticles of large
sizes. To capture the size dependence only we exlude sizes $N\sim55$
($N=54,55,$ and $56$) from the fit, because these nanoparticls acheive
additional stability by forming highly symmetric icosahedra (see
Fig. \ref{label_figure_10}). We obtain $a_{free}= -4.29\pm 0.01$ eV,
$b_{free}=2.42\pm 0.03$ eV and $a_{supp}=-4.30\pm 0.01$ eV,
$b_{supp}=2.34\pm 0.02$ eV for free and supported clusters,
respectively.  The two values of $a$ are very close to the bulk fcc
cohesive energy of -4.29 eV/atom; the two values of $b$ slightly
differ because of the reduced total energy cost to create and modify
surface for the supported cluster \cite{Ding3}.

Figures \ref{label_figure_11} and \ref{label_figure_12} illustrate the
differences between the calculated and the fitted energies for free
and supported clusters. The prominent negative peaks in
Fig. \ref{label_figure_11} for sizes $N=55,61,64,71,75,$ and $77$
correspond to the magic size structures in our sequence. The evolution
of the nanoparticle configuration with size is depicted in the same
picture.
By comparing Figures \ref{label_figure_11} with \ref{label_figure_12}
we observe that the attractive substrate has little or no effect on
the internal structure of small clusters ($N\lesssim70$). Due to the
spherical (or nearly-spherical) arrangements, such small
icosahedra are not significantly deformed by attractive substrates,
unless the adsorption potentials is comparable to the internal
atomic binding energy of the cluster.
Big supported particles behaves differently. In fact, as the size
increases beyond $N\gtrsim70$, alumina-supported clusters form fcc
arrangements (Fig. \ref{label_figure_12}) with the \{111\} 
planes at the particle/substrate
interface. Elsewhere the particles have simple close-packed facets or local
arrangements of them. 

Big unsupported particles are different than the supported ones.
Unsupported clusters of sizes between $N\sim70$ and $N\sim200$ have
decahedra or icosahedra arrangements (multi-twinned structures
\cite{Ino1,Ino2}). Decahedra are more frequent for medium size
particles ($N\sim$70 to $\sim$100) while icosahedra appear more often
for bigger sizes ($N\sim$100 to $\sim$200).  Figure
\ref{label_figure_11} shows three particular examples of decahedra
clusters with $N=71,75$, and 77.  As the size exceeds $N\sim200$, the
particles approach the bulk configuration and the fcc and icosahedra
structures tend to become degenerate. For instance, the cluster at
$N=500$ is a good example of an fcc particle with appropriate
close-packed faceting and minimal surface-edge reorganization.


The \polyfcctrans\ that occurs at lower sizes for supported clusters
can be explained in terms of surface/interface and elastic strain
energies \cite{Ino3,Johnston}. In a groundstate configuration, the
former (latter) energy dominates over the other for a small (big)
particle. Particles with radii of $r<r_0$ ($r_0$ is the critical size
for the polyhedron-fcc transition \cite{r0def}) tend to form
polyhedral structures (icosahedra or decahedra) to minimize the
surface/interface energy, while clusters with $r>r_0$ will prefer the
bulk configuration (fcc) to minimize the elastic strain
energy\cite{Johnston}. This is true for free and supported clusters.
However, for supported particles, the fcc configuration with flat
facets (along the directions of minimum surface energy given by the
Wulff plots \cite{Wulff,Herring}) has a larger contact area with the
substrate than the polyhedra do.  Therefore, the attractive substrate
interaction reduces the surface/interface energy of fcc the most,
resulting in a lower critical $r_0$ (even though the overall surface
area of the fcc structures could be larger than that of the
icosahedra).

Our simulations reveal that the substrate attractive
interaction tends to preserve the bulk structure for small
particles. This demonstrates that although the Fe-Al$_2$O$_3$
interaction is relatively weak, it is able to influence the balance
between the bulk and surface energies in competing configurations and
utlimately determine the particle's ground state. One can also expect
to observe this effect in simulations with different models of the
Fe-Fe and Fe-Al$_2$O$_3$ interactions, as long as their relative
strength is comparable with ours.

\section{Conclusions} 
\label{section.conclusions}

In this paper we have investigated the behavior of free and 
alumina-supported Fe-C nanoparticles.
We observe interesting phenomena that can be attributed to the presence of the substrate.
The main results of the present study can be summarized as follows: 

(i) 
The total Fe binding can be conveniently decomposed into
independent Fe-Fe and Fe-Al$_2$O$_3$ parts: according
to our {\it ab initio} calculations the two differ by about an order
of magnitude. Moreover, the corrugation of the Fe-Al$_2$O$_3$
interaction is much smaller than the average adsorption energy. This
allows us to parameterize the Fe-Al$_2$O$_3$ interaction as a
Morse potential, which includes the deformation energy of the
substrate. 

(ii) The thermal behavior of pure-Fe particles is simulated with
classical MD techniques. We observe the reduction of melting
temperature as a function of the diameter, in agreement with the
Gibbs-Thomson law. We also show that supported particles have
higher melting points than the unsupported ones.

(iii)
We calculate the liquidi on the phase diagrams 
for a range of Fe-C nanoparticles ($d\sim 1.1-1.6$ nm) 
and show that they are characterized by the presence of eutectic points 
in which the eutectic concentration depends on the size of the particles.
These phenomena may have important effects on the growth of carbon nanotubes by CVD, as discussed in Section III C.

(iv)
We find that the optimized configurations of very small
pure Fe clusters (down to $N_{Fe}\sim 50$) have icosahedron or decahedron structures.
Bigger clusters tend to have close-packed configurations.
We show that the size at which the cluster undergoes the
polyhedron$\rightarrow$close-packed transition depends on the substrate.
In particular, Al$_2$O$_3$ supported Fe clusters 
have $N_{tr}^{Al_2O_3} \sim 70 $ while free cluster 
have $N_{tr}^{free} \sim  200-500$.

The thermodynamic study of nanoparticles will be extended with future {\it ab initio} 
characterizations carried out in this laboratory, and with experimental
investigations performed by our collaborators.
This work stimulates more comprehensive studies of the role of
substrates on the thermodynamics of small particles,
to understand the fundamental factors controlling 
the catalytic properties of small clusters.

We wish to acknowledge helpful discussions with 
T. Tokune, 
E. Mora, 
H. Duan, 
A. Ros\'en, 
F. Ding,
A. Ferrari,
L. Boeri,
and 
M. Cole. 
The authors are grateful for time allocated on the Swedish National Supercomputing facilities. 
This research was supported by Honda Research Institute USA, Inc.

%




\begin{thebibliography}{99}
\bibitem{Parm_note}
The highly disperse unstable solutions are oversaturated with respect to the thermodynamic equilibrium
``graphite $\leftrightarrow$ solution of carbon in the metal'' which is kinetically slow to reach starting 
from the mixture ``amorphous carbon $\leftrightarrow$ oversaturated solution of carbon''. 
The stability of the disperse solutions should be compared to the latter mixture and not to the
thermodynamic equilibrium (see Refs. \cite{Kriv} and \cite{Parm}).
\bibitem{Ding_note}
A. C. Ferrari and S. Curtarolo, Private comunications regarding reference number \#19 of 
Ref. \cite{Hofmann1}, which instead should point to Ref. \cite{Ding1}. 
In the latter work, the eutectic temperature is studied only for one particle size (Fig. 3 of Ref. \cite{Ding1})
and not in function of size and substrate as in our manuscript.
\bibitem{LSFit} 
We use the least square method to determine the four
parameters of the fitting function ($\cconc_{eut}$, $T_{eut}$, and the
slopes of the two straight lines).
\bibitem{r0def}
$E_{total} = E_{cohesive} + A r^2 + B r^3$, where $A$ and $B$ are the
surface and elastic strain energy coefficients, respectively,
and $r$ is the average radius of the particle.  Assuming a constant
cohesive energy for clusters of the same size, the
polyhedron$\rightarrow$fcc transition occurs at a critical radius
$r_0 =(A_{fcc}-A_{poly})/(B_{poly}-B_{fcc})$ because of the inherent
structural relationships $A_{poly}<A_{fcc}$ and $B_{poly}>B_{fcc}$
\cite{Ino3,Johnston}.
\end{thebibliography}
\end{document}